# Origin and Propagation of Ultra-High Energy Cosmic Rays

G.A. Medina Tanco, **E.M. de Gouveia Dal Pino** & J.E. Horvath

*University of São Paulo, Instituto Astronômico e Geofísico, Brasil*

## Introduction

With the advent of large detectors, the existence of cosmic ray particles up to the so called ultra-high energy limit ($>10^{20}$ eV) is now beyond any doubt. The existence of such cosmic particles imposes a challenge for the comprehension of their sources and nature. On one side, particles with such high energies are difficult to be produced by *any* astrophysical source. On the other side, the interaction of these particles with photons of the cosmic microwave background (CMB) cause substantial losses of energy which constraint the maximum distances that the particles are able to travel from the sources to the detectors.

Aiming to help to elucidate the problem of UHECR source identification, we have first performed 3-D simulations of particle trajectories propagated through the stochastic intergalactic and an extended Galactic halo magnetic fields. Going further in this investigation, we have then performed simulations of proton and Fe nuclei through the spiral Galactic magnetic field (GMF) and built full-sky maps of their arrival direction distribution in both the detector (after deflection in the GMF) and outside the galaxy. In the sections below we summarize the main results of these investigations (see also [1, 2, 3] for details).

## 2. UHECR propagation through the IGM and Galactic halo magnetic fields

To explicitly follow the trajectories UHECR through the intergalactic medium (IGM) and the galactic halo, we have assumed that the magnetic field in the propagation region is uniform on scales smaller than its correlation length $L_c$ [4]. The 3-D space is divided into domains randomly generated out of a normal distribution of average length-scale $L_c$ and 10% standard deviation. The magnetic field is assumed to be homogeneous inside each domain and randomly oriented with respect to the field in adjacent domains. Particles are injected into the system at different energies and pitch angles with respect to the line of sight. Since we are interested only in the UHE regime, we have only considered the dominant loss mechanism of photomeson production [4]. Several numerical experiments have been performed by injecting

~8 $10^5$ particles. Fig. 1 depicts the arrival energy of UHE protons injected with monochromatic energies ($E_{inj}$ = $10^{21}$ to $10^{23}$ eV). From Fig. 1, we find that the progenitor sources cannot be more distant than ~50 Mpc, otherwise the injection energy must be unreasonably high. We have also evaluated the time delays between the UHECR protons and the gamma-rays that they produce and found that they are too large to match the claimed association between the UHECR and observed events of gamma-ray bursts (GRBs) [5].

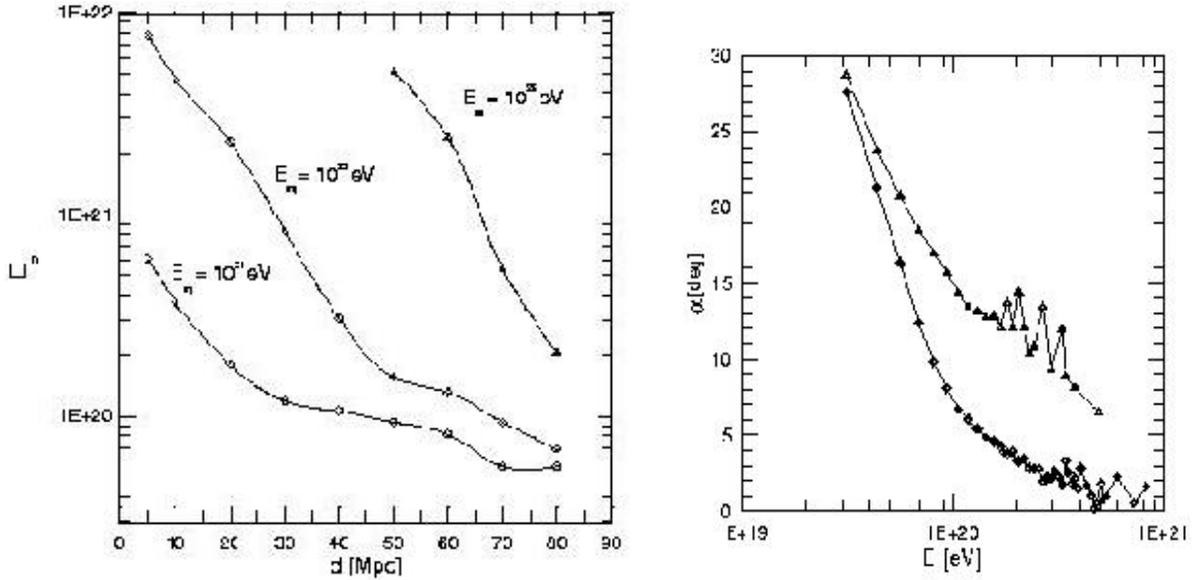

**Figure 1 (left)**: Arrival energy of UHE protons E as a function of the distance to the source d for several values of the (monochromatic) injection energy $E_{inj}$.
**Figure 2 (right)**: Arrival angle α of a proton of energy E for an extragalactic source at d, =50 Mpc. The curves are given for $B_{IGM}$ = $10^{-9}$ G and $L_c$ = 1 Mpc without considering the effects of $B_H$ (lower curve) and, for the same values of the IGM, with the inclusion of a maximally magnetized halo having $B_H$ = $10^{-6}$ G, $L_c$ = 1.5 kpc and halo radius $R_H$ = 100 kpc (upper curve).

Fig. 2 depicts the arrival direction angles for protons injected with a power-law spectrum [N(E) ∝ $E^{-2}$] from an extragalactic source at a distance d = 50 Mpc. It can be appreciated from the lower curve (which considers only the influence of the IGM magnetic field) that the IGM by itself is not enough to produce considerable deflection for protons of E ≥ $10^{20}$ eV. Therefore, within an error circle of at most 8°, the particles point to their sources. These error circles reduce to a mere 2° for the highest energy detected events. Thus, an association with sources lying in the supergalactic plane [6] seems to be unlikely. On the other hand, with the inclusion of a strongly magnetized galactic halo (upper curve), the deflection angles become larger and in this case a source located in the supergalactic plane can not be excluded. We emphasize, however, that this is a rather extreme and unlikely situation.

### 3. UHECR propagation through the spiral magnetic filed of the Galaxy

The results above indicate that as the particles reach the Galaxy, they seem to strongly point to their sources (with deflection angles no larger than ~ 8°). Now, let us turn to the propagation of the UHECR through the Galactic magnetic field (GMF) itself. We have assumed an axi-symmetric, spiral field without reversals extending to galactocentric distances of 20 kpc in all radial directions, and with an even (quadrupole type) parity in the perpendicular direction (z) to the galactic plane [7]. The field is taken to be constant at the central region of the Galaxy (with a value 6.4 µ G within r = 4 kpc). The radial dependence is consistent with field strengths inferred from pulsar rotation measures. Also, a magnetic field exponentially decaying with height from the galactic plane has been generally assumed. In order to trace the particle trajectories through the GMF, we have applied the *reversibility principle* [e.g. 7] by backtracking antiprotons. An antiproton injected at the Earth in a certain direction will follow the same trajectory as a proton arriving at the same direction at the Earth. Assuming an isotropic distribution, we have injected antiprotons at different galactic longitudes (b) and latitudes (l) at the Earth, and followed their propagation through the GMF until they reached the galactic halo. In this way, we could determine the particle position when it arrived the Galaxy, before its direction was altered by the GMF.

We have found that the presence of a large scale galactic magnetic field does not generally affect the arrival directions of the protons, as expected, although the inclusion of constant z-component ($B_z$ = 0.3 µ G) may cause significant deflection of the lower energy protons (E = 4 $10^{19}$ eV). Error boxes ≥ 5° are expected in this case. On the other hand, in the case of the heavy nuclei (Fe), the arrival direction of the particles is strongly dependent on the coordinates of the particle source. Fig. 3 indicates that the deflection in this case may be high enough as to make extremely difficult any identification of the sources unless the real magnetic field configuration is known. Moreover, not every incoming particle direction is *allowed* between a given source and the detector. This generates sky patches that are virtually unobservable from the Earth. We also note from Fig. 3 that observed UHE events of Yakutsk, Fly's Eye, and Akeno casually come from sites for which the deflection caused by the assumed magnetic field is not significant.

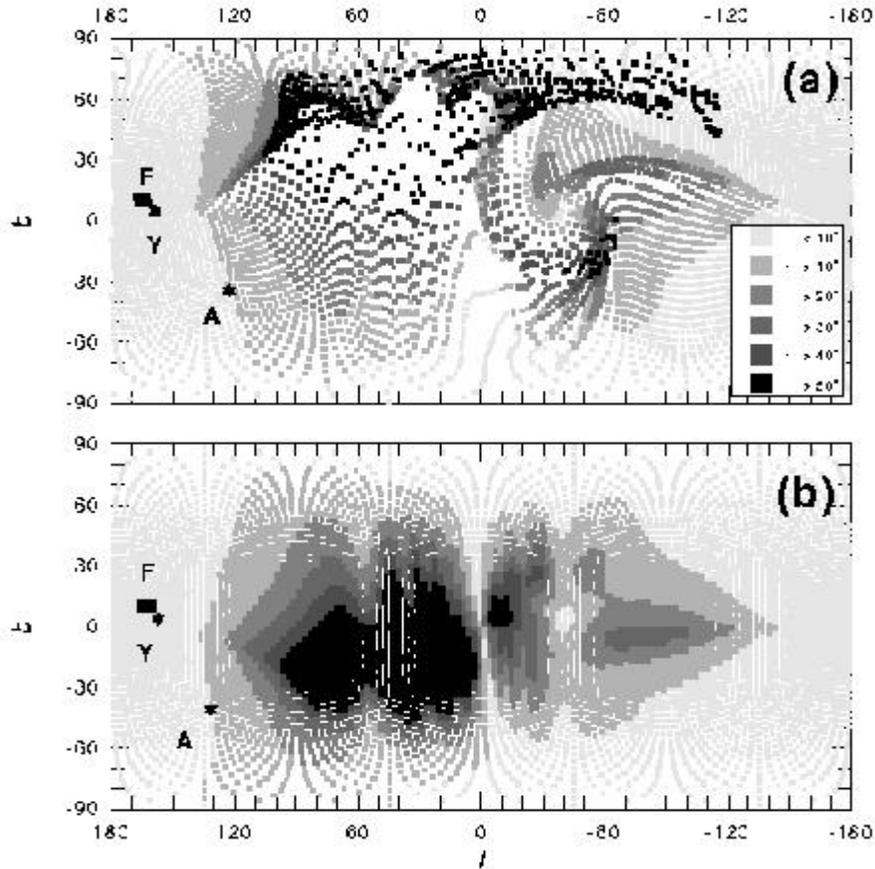

**Figure 3**: Sky map in galactic coordinates (l,b) of the distribution of Fe nuclei of energy 2.5 $10^{20}$ eV. The UHECR positions are distributed in different shades of gray according to their deflection angles due to passage through the GMF. (a) Depicts the arrival directions of Fe nuclei at the galactic halo border (before deflection by the GMF)); and (b) depicts the particle arrival directions at the detector (after deflection by the GMF). Note that the observed UHE events Akeno, Fly's Eye, and Yakutsk (marked with symbols) do not reveal strong deflections from their original positions.